\def\bfseries{\fontseries\bfdefault\selectfont\boldmath}
\def\itshape{\fontshape\itdefault\selectfont\let\mathrm=\mathit}

\documentclass[twocolumn,showpacs,preprintnumbers,amsmath,amssymb]{revtex4}
\pdfoutput=1
\usepackage{multirow}
\usepackage{atlasphysics}
\usepackage{subfigure}
\usepackage{graphicx}
\usepackage{dcolumn} 
\usepackage[hyperindex,breaklinks]{hyperref} 
\usepackage{preprintcover} 
\PreprintCoverPaperTitle{Search for resonant top plus jet production in $\ttbar$ + jets events with the 
ATLAS detector in $pp$ collisions at $\sqrt{s}$ = 7\,\TeV} 
\PreprintIdNumber{CERN-PH-EP-2012-219} 
\PreprintCoverAbstract{This paper presents a search for a new heavy particle produced in association with a top or antitop quark. 
Two models in which the new heavy particle is a color singlet or a color triplet are considered, decaying respectively to $\bar{t}q$ or $tq$, 
leading to a resonance within the $\ttbar$ + jets signature. 
The full 2011 ATLAS $pp$ collision dataset from the LHC (4.7~fb$^{-1}$) is used to search for $\ttbar$ events produced
in association with jets, in which one of the $W$ bosons from the top quarks decays leptonically and the
other decays hadronically. The data are consistent with the Standard Model expectation, and a new particle 
with mass below 430~\gev for both $W'$ boson and color triplet models is excluded at 95\% confidence level, assuming unit right-handed coupling.}  
\PreprintJournalName{PRD Rapid Communications}
\pacs{14.80.-j}

\begin{document}
\title{Search for resonant top plus jet production in $\ttbar$ + jets events with the 
ATLAS detector in $pp$ collisions at $\sqrt{s}$ = 7\,\TeV}

\begin{abstract}

This paper presents a search for a new heavy particle produced in association with a top or antitop quark. Two models in which the new heavy particle is a color singlet or a color triplet are considered, decaying respectively to $\bar{t}q$ or $tq$, leading to a resonance within the $\ttbar$ + jets signature. 
The full 2011 ATLAS $pp$ collision dataset from the LHC (4.7~fb$^{-1}$) is used to search for $\ttbar$ events produced
in association with jets, in which one of the $W$ bosons from the top quarks decays leptonically and the
other decays hadronically. The data are consistent with the Standard Model expectation, and a new particle 
with mass below 430~\gev for both $W'$ boson and color triplet models is excluded at 95\% confidence level, assuming unit right-handed coupling.

\end{abstract}

\maketitle

\begin{figure}[h!tbp]
\centering
\subfigure[]{\includegraphics[width=0.46\columnwidth]{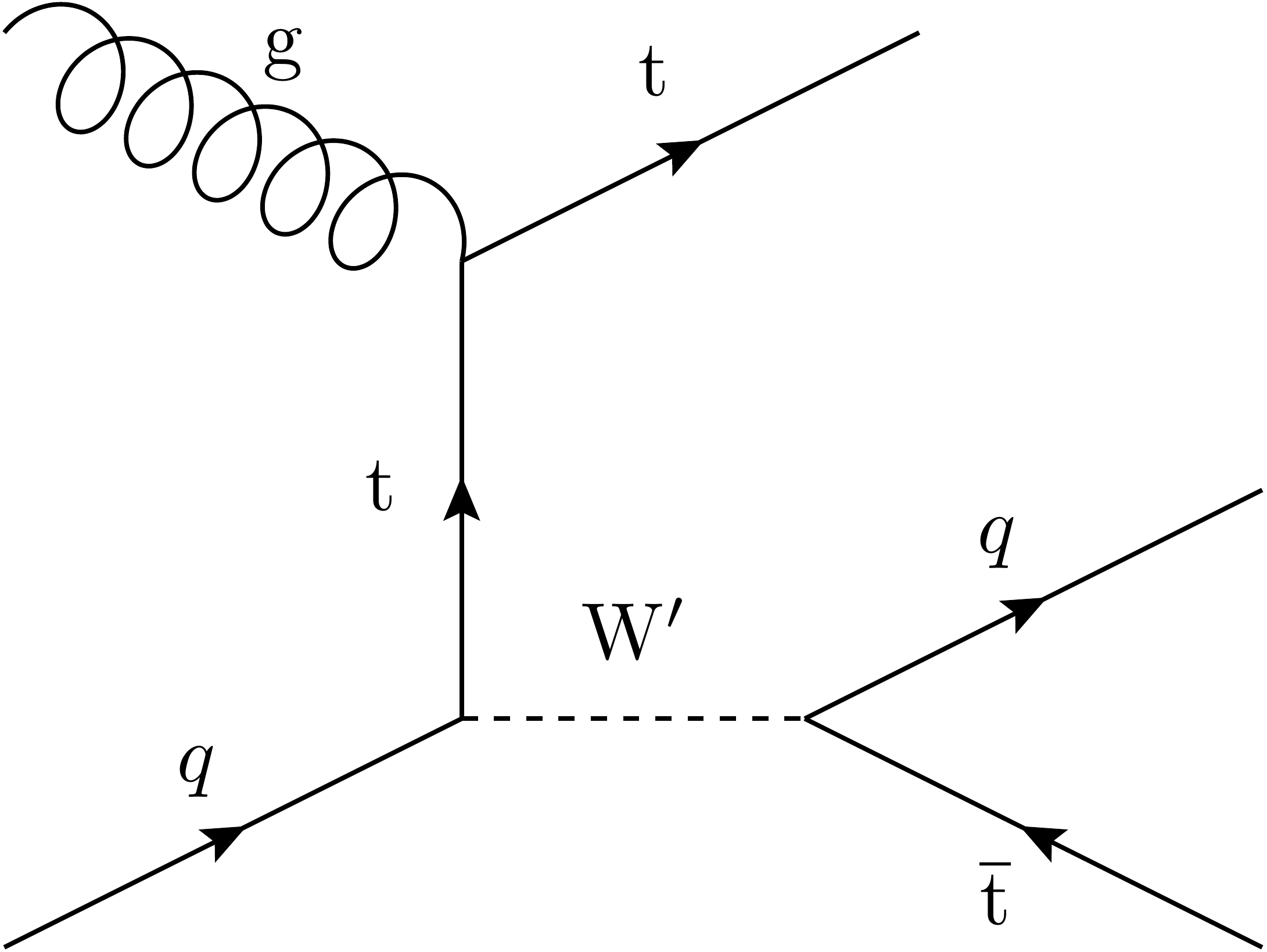}}
~~~~~~\subfigure[]{\includegraphics[width=0.46\columnwidth]{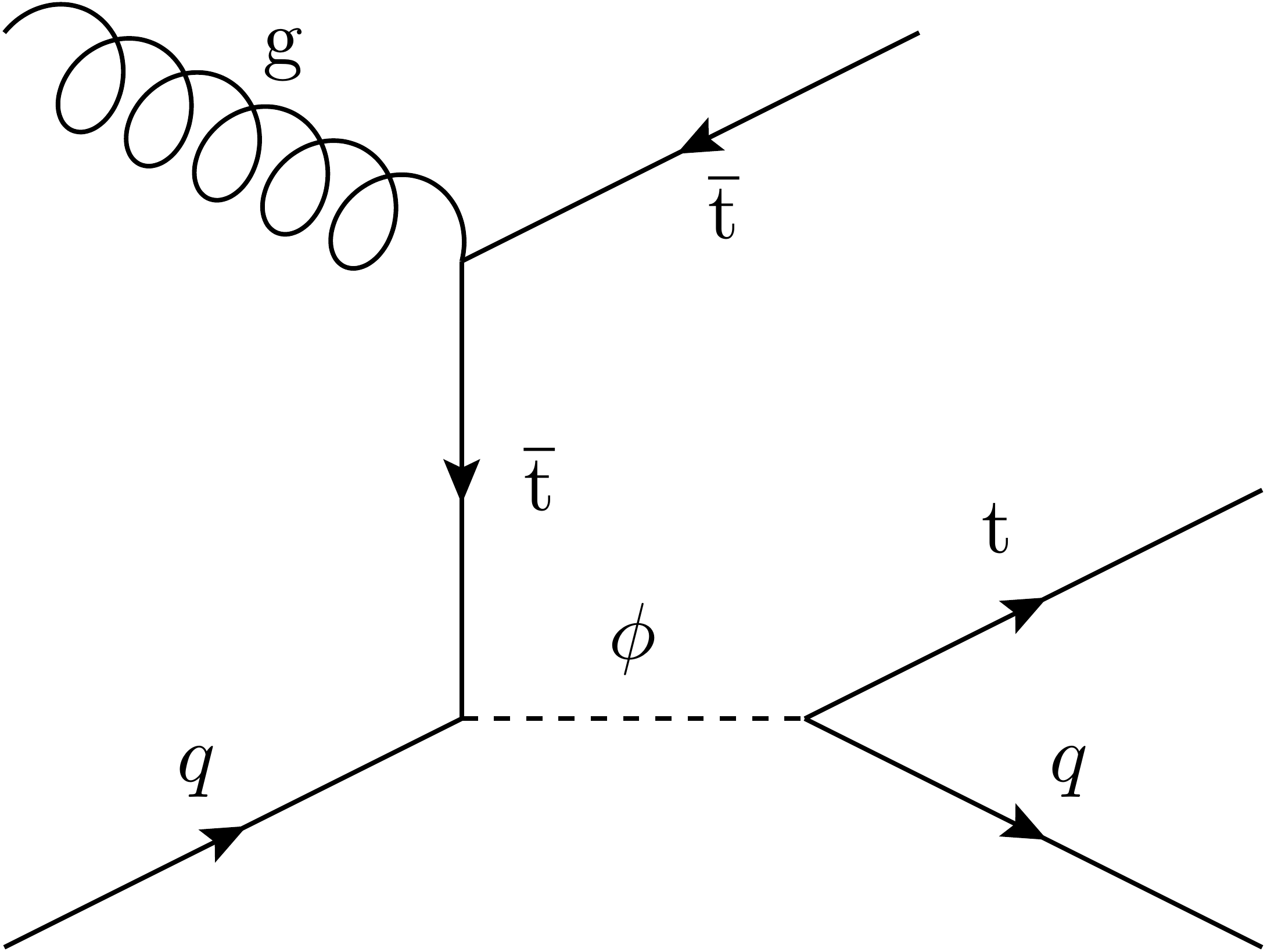}}
   \caption{Example production and decay Feynman diagrams for the (a) $W'$ and (b) $\phi$ models.
  \label{fig:feynman}
  }
\end{figure}

In the past few decades, remarkable agreement has been shown
between measurements in particle physics and the predictions of the Standard Model (SM). The top quark sector is one important place
to look for deviations from the SM, 
as the large top quark mass suggests that it may play a special
role in electroweak symmetry breaking. 
The recent top quark forward-backward asymmetry measurements from
the Tevatron experiments~\cite{cdfafb,d0afb} are in marginal agreement with
SM expectations. A non-SM explanation could come from a possible
top-flavor-violating process~\cite{TFVResonances,TFVResonances2,TFVResonances3}.
In these models, a new heavy particle $R$ would be produced at the
LHC in association with a top or anti-top quark. Figure~\ref{fig:feynman} shows representative production diagrams
for these new particles, for the cases of $R=W'$ or $R=\phi$ (see below).
As shown in Ref.~\cite{NLO-wprime}, the production mechanism in $pp$ collisions mainly involves
quarks rather than anti-quarks at $\sqrt{s}$=7\,TeV,
even for relatively low mass particles.

The larger number of quarks relative to anti-quarks produced in the initial state at the LHC
leads to a resonance $R$ that decays predominantly to either the $t$+jet or $\bar{t}$+jet final state,
where baryon number conservation restricts the models that are available.
Two models that can give rise to these final states are a color singlet resonance ($W'$) mostly in the $\bar{t}q$ system, and a di-quark color triplet model with a resonance ($\phi$) in the $tq$ system.
In both cases a $\ttbar$+jet final state is produced, but a
peak will be present in only one of the $t$+jet or $\bar{t}$+jet
invariant mass distributions.
The new resonances are assumed not to be self-conjugate, which
makes searches for same-sign top quarks insensitive to them~\cite{samesign1,samesign2,samesign3}, and to have only right-handed couplings.
The $t$ or $\bar{t}$ then decays to $W^+b$ or $W^-\bar{b}$, respectively. 
This paper considers the decay signature of events in which
one $W$ boson decays leptonically (to an electron or muon, plus neutrino final state) and the 
other $W$ boson decays hadronically.  
The first direct search for such particles was performed at CDF~\cite{cdfsearch}, 
which excluded color triplet resonances with masses below 200\,\gev
and $W'$ resonances with masses below 300\,\gev, for particles with unit right-handed coupling $(g_R)$ to $tq$. 
As is done in this paper, CDF used the formalism in Ref.~\cite{TFVResonances} to define $g_R$.
CMS recently performed a search that excluded a new $W'$ with a mass less than 840~\gev~\cite{cmssearch} for
particles with $g_R=2$~\cite{cmsfootnote}.

The analysis presented here uses the full ATLAS 7\,\tev $pp$ collision dataset collected in 2011, 
corresponding to 4.7$\pm$0.2~\ifb of integrated luminosity~\cite{b-lumi,b-lumi2} delivered by the LHC. 
ATLAS~\cite{atlas} is a multi-purpose particle physics detector with 
cylindrical geometry~\cite{atlasfootnote}.
The inner detector (ID) system consists of a 
high-granularity silicon pixel detector and a silicon micro-strip detector, as well as 
a transition radiation straw-tube tracker. The ID is immersed in a 2\,T axial magnetic field, 
and provides charged particle tracking in the range $\vert\eta\vert<$~2.5. 
Surrounding the ID, electromagnetic calorimetry is provided by barrel and endcap 
liquid-argon (LAr)/lead
accordion calorimeters and LAr/copper sampling calorimeters in the forward region. 
Hadronic calorimetry is provided in the barrel by a steel/scintillator tile sampling
calorimeter, and in the endcaps and forward region by LAr/copper and LAr/tungsten sampling calorimeters, respectively. 
The muon spectrometer (MS) comprises separate trigger and high-precision tracking chambers measuring
the deflection of muons in a magnetic field with a bending power of 2--8\,Tm,
generated by three superconducting air-core toroid systems. 
A three-level trigger system is used to select interesting events. The level-1 trigger is implemented in
hardware and uses a subset of detector information to reduce the event rate to a design value of at most
75\,kHz. This is followed by two software-based trigger levels, level-2 and the event filter, which together
reduce the event rate to $\sim 300$\,Hz.

Events with an electron (muon) are required to have passed an electron (muon) 
trigger with a threshold of transverse energy $E_{\mathrm{T}}>20$\,\gev (transverse momentum $p_{\mathrm{T}}>18$\,\gev), 
ensuring that the trigger is fully efficient for the off-line selection discussed below.
Electrons reconstructed offline are required to have a shower shape in the
electromagnetic calorimeter consistent with expectation, as well as a good quality track pointing to the cluster
in the calorimeter. Candidate electrons with $\et>$ 25\,\gev are required to pass the ``tight'' 
electron quality criteria~\cite{ref-electron}, 
to fall inside a well-instrumented region of the detector ($|\eta|<$ 2.47, excluding 1.37 $<|\eta|<$ 1.52), 
and to be isolated from other objects in the event.
Muons with transverse momentum $\pt>$ 20\,\gev are required to pass 
muon quality criteria~\cite{b-mcp}, to be well measured in both the ID and the muon spectrometer, to fall within
$|\eta|<$ 2.5, and to be isolated
from other objects in the event. 

Jets are reconstructed in the calorimeter using the anti-$k_t$~\cite{b-akt} algorithm 
with a radius parameter of 0.4. Jets are required to satisfy $\pt > $ 25\,\gev and $|\eta| <$ 2.5. Events
with jets arising from electronic noise bursts and beam backgrounds are rejected~\cite{b-cleaning}. Jets are 
calibrated to the hadronic energy scale using $\pt$- and $\eta$-dependent corrections derived from simulation, as well as from test-beam and collision data~\cite{jets}. Jets from the decay of heavy flavor hadrons are selected by 
a multivariate $b$-tagging algorithm~\cite{b-tagging} at an operating point with 70\% efficiency 
for $b$-jets and a mistag rate for light quark jets of less than 1\% in simulated $\ttbar$ events.
Neutrinos are inferred from the magnitude of the missing transverse momentum $(E_{\mathrm{T}}^{\mathrm{miss}})$ in the event~\cite{b-MET}. 

The signal region for this analysis is defined by requiring exactly one charged lepton and five or more jets, including at least
one $b$-tagged jet. To select events with a leptonically decaying $W$ boson, events are required 
to have $\met>30$\,\gev ($\met>20$\,\gev) in the electron (muon) channel. Additionally, the event must have a transverse
mass of the leptonically-decaying $W$ boson $m^W_{\mathrm{T}}>30$~\gev in the electron channel, or scalar sum
$\met+m^W_{\mathrm{T}}>60$\,\gev in the muon channel~\cite{b-toppaper}. Here, $(m^W_{\mathrm{T}})^2 = 2\met E_{\mathrm{T}}^{\ell}(1-\cos{\phi})$,
where $E_{\mathrm{T}}^{\ell}$ is the magnitude of the transverse momentum of the lepton, and $\phi$ is the angle
between the lepton and the missing transverse momentum in the event.

A variety of Monte Carlo generators are used to study and estimate
backgrounds. The generated events are processed through full detector simulation~\cite{simulation}, based on 
GEANT4~\cite{b-geant4}, and include the effect of multiple $pp$ interactions per bunch crossing. To predict the event yield, the simulation
is given an event-by-event weight such that the distribution of the number of $pp$ collisions
matches that in data.

The $\ttbar$ background is modeled with MC@NLO v4.01~\cite{b-mcnlo} interfaced to HERWIG v6.520~\cite{b-herwig} and JIMMY v4.31~\cite{b-jimmy}.
An additional $\ttbar$ sample modeled with MC@NLO interfaced to PYTHIA v6.425~\cite{b-pythia}
is used to study potential systematic uncertainties.
Other $\ttbar$ samples use POWHEG~\cite{b-powheg} interfaced either to PYTHIA or HERWIG, as well as AcerMC v3.8~\cite{acer}. 
The background from the production of single $W$ bosons in association with extra jets is modeled by
the ALPGEN v2.13~\cite{b-alpgen} generator interfaced to HERWIG. The MLM matching scheme~\cite{b-mlm} is used
to form inclusive $W$ boson + jets samples such that overlapping events produced in both the hard scatter and 
parton showering are removed. In addition, the heavy flavor contributions are reweighted 
using the data-driven procedures of Ref.~\cite{whf} using the full 2011 LHC dataset. Diboson events are generated using HERWIG. Single-top-quark events
are modeled by MC@NLO, interfaced with HERWIG for the parton showering, in the $s$-channel and $Wt$ channel,
and by AcerMC v3.8 in the $t$-channel. The small background in which multi-jet processes 
are misidentified as prompt leptons is modeled from a data-driven matrix method~\cite{matrixmethod}.
In determining the expected event yields, the $\ttbar$ cross section is normalized to
approximate next-to-next-to-leading-order QCD calculations of $167^{+17}_{-18}$~pb for a top quark mass of 172.5\,GeV~\cite{tt-xsec,tt-xsec2}, and
the total $W$+jets background is normalized to inclusive next-to-next-to-leading-order 
predictions~\cite{w-xsec}. Signal events are produced, for a range of
$W'$ and $\phi$ masses, with MadGraph v5.1.3.16~\cite{madgraph} and
interfaced to PYTHIA v6.425. 
Next-to-leading-order (NLO) cross sections are used for the predicted $W'$ boson signal normalization~\cite{NLO-wprime}, 
and leading-order (LO) cross sections using MSTW2008 are 
used for the $\phi$-resonance normalization~\cite{TFVResonances}.

Events are reconstructed with a kinematic fitting algorithm
that utilizes knowledge of the over-constrained $\ttbar$ system to assign jets to partons. 
In the fit, the two top quark masses are each constrained at the particle level to 172.5\,\gev by a penalty 
in the likelihood, computed from variations from
this nominal value and the natural top quark width of 1.5\,\gev.
The two $W$ boson masses are similarly constrained to 80.4\,\gev 
within a width of 2.1\,\gev. This allows the
$z$-component of the momentum of the neutrino from the leptonically decaying $W$ boson to be computed. 
Both solutions from the quadratic ambiguity of this computation are tested
when computing the likelihood.
Charged lepton, neutrino and jet four-momenta are constrained in the fit by
resolution transfer functions derived from simulated $\ttbar$ events that relate 
the measured momenta in the detector to true particle momenta. 
The full shapes of these transfer functions are used in the likelihood computation.
All assignments of any four jets to partons from the $\ttbar$ decay
are tested and the assignment with the largest likelihood output for the $\ttbar$ hypothesis is selected.
After the assignment is selected, the originally measured
jet and lepton momenta and \met are used.
The remaining jets not associated with the $\ttbar$ partons are included to form $m_{tj}$ and 
$m_{\bar{t}j}$ masses, where the charge of the lepton is used to
infer which is the top candidate and which is the anti-top candidate.
All combinations of extra jets with the top and anti-top quark candidates are considered, and 
the pairings that give the largest $m_{tj}$ and $m_{\bar{t}j}$ masses are used. In this way, 
the same extra jet can (but does not necessarily have to) be used to form $m_{tj}$ and $m_{\bar{t}j}$. 
These two masses are used as observables for the search.

Several control regions are used to ensure good modeling and understanding of the backgrounds before
the signal region is examined.
The preselection control region requires at least four jets, but does not require a b-tag.
The dominant $\ttbar$ background is tested in a control region with
exactly four jets (including at least one $b$-tagged jet). The rejection of events with
more than four jets reduces signal contamination.  
A second $\ttbar$ control region is defined by events with exactly four jets with $\pt$ above 25\,\gev, one 
of which must be $b$-tagged, and exactly one additional jet with $\pt$ between 20\,\gev and 25\,\gev. Signal contamination is 
further reduced by requiring that the $\Delta R\equiv \sqrt{(\Delta\eta)^2+(\Delta\phi)^2}$ 
between the fifth jet and both the reconstructed top 
and anti-top quarks is greater than $\pi/2$. Figure~\ref{fig:cr-ttbar} shows distributions 
in the two $t\bar{t}$ control regions, where good agreement is observed between data and the prediction.
The second major background, production of single $W$ bosons in association with extra jets, is
tested in a control region with five or more jets, vetoing events with $b$-tagged jets. The requirement of zero $b$-tagged jets reduces both
signal and $\ttbar$ contamination. 
The distribution in Figure~\ref{fig:cr-w} shows good agreement between
data and the prediction within uncertainties. Table~\ref{tab:CR} summarizes the 
expected and observed yields in the control regions.

\begin{figure}[h!tbp]
\subfigure[]{\includegraphics[width=\columnwidth]{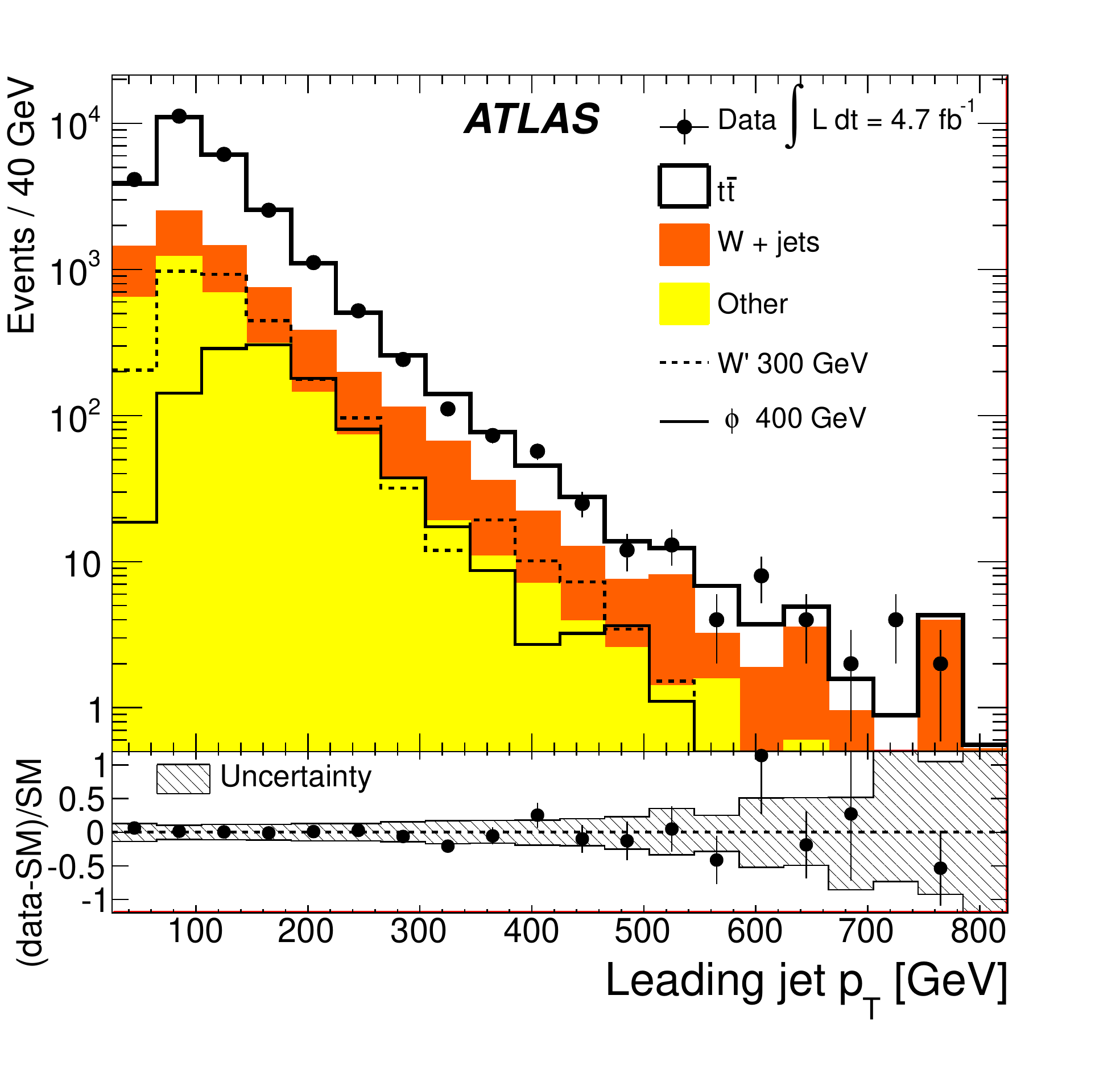}}
\subfigure[]{\includegraphics[width=\columnwidth]{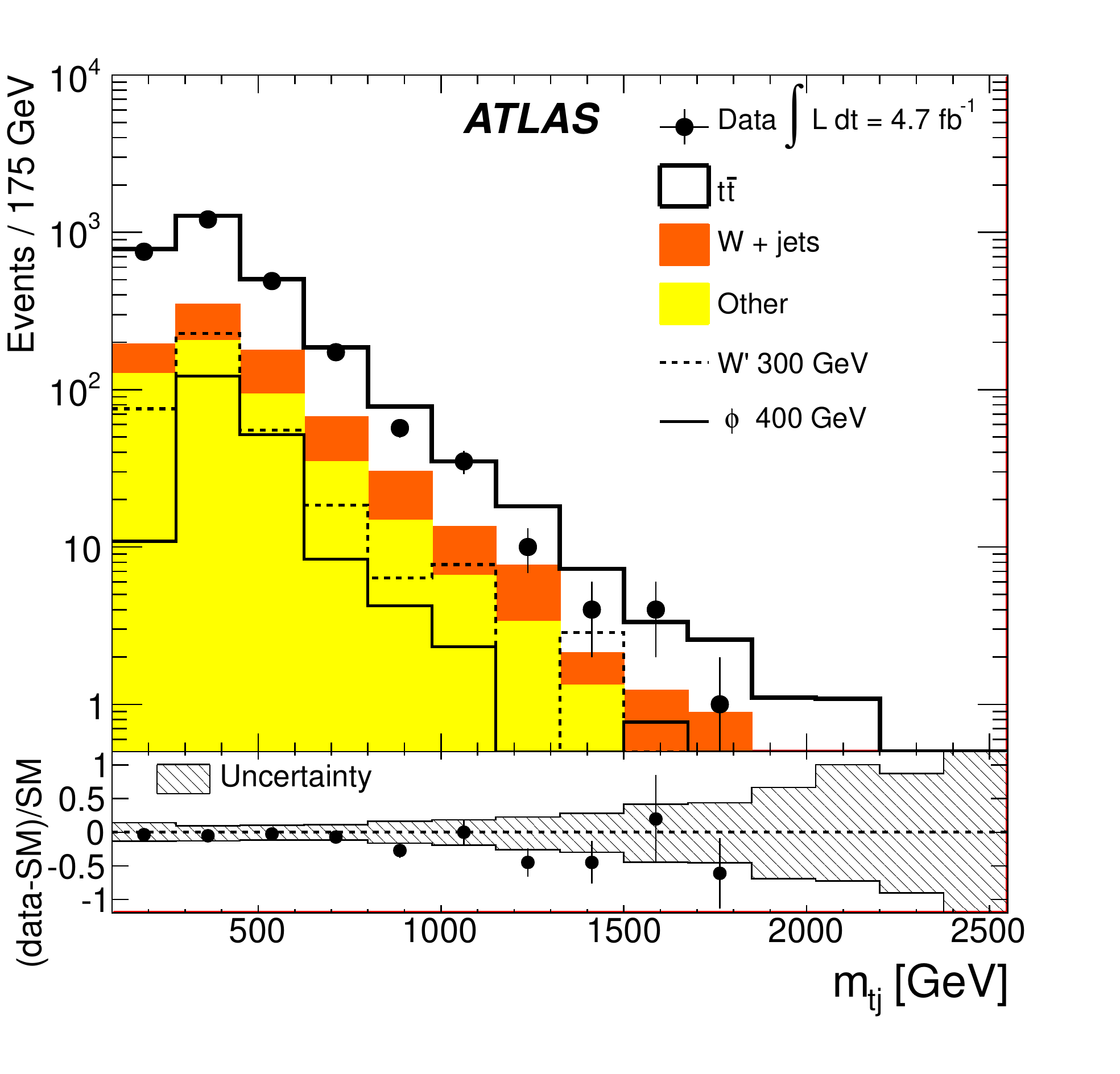}}
   \caption{The leading jet $\pt$ in the four-jet $\ttbar$ control region (a), and $m_{tj}$ in 
the five-jet $\ttbar$ control region (b). The example signal-only distributions are overlaid for comparison,
where unit coupling for the new physics process is assumed.
The total uncertainty shown on the ratio includes both statistical and systematic effects. The  ``other'' 
background category includes single top production, diboson production and multi-jet events.
  \label{fig:cr-ttbar}
  }
\end{figure}

\begin{figure}[h!tbp]
\includegraphics[width=\columnwidth]{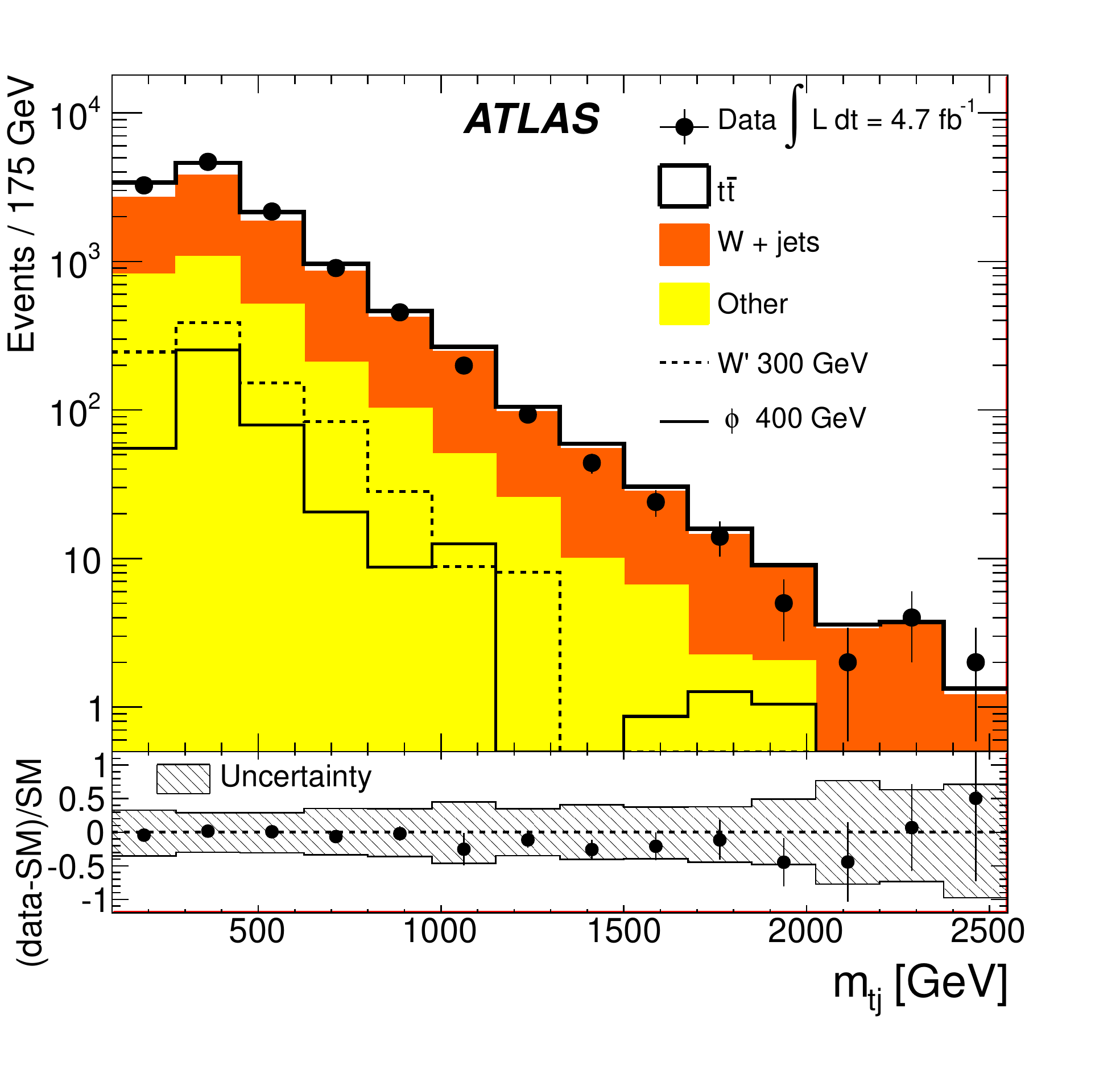}
   \caption{Expected and observed distribution of $m_{tj}$ in the $W$+jets control region. 
The example signal-only distributions are overlaid for comparison,
where unit coupling for the new physics process is assumed.
The total uncertainty shown on the ratio includes both statistical and systematic effects. The ``other'' background 
category includes single top production, diboson production and multi-jet events.
\label{fig:cr-w}
  }
\end{figure}

\begin{table*} 
\caption{Expected and observed yields in the four control regions (CR). 
Total refers to the total expected background, including $\ttbar$, $W$+jets and the other 
smaller backgrounds: single top production, diboson production and multi-jet events.
The last two lines show the expected number of events for two benchmark signal samples
in each of these control regions.  The errors include all systematic uncertainties.}
\label{tab:CR}
\begin{tabular}{l|D{,}{\,\,}{20}D{,}{\,\,}{20}D{,}{\,\,}{20}D{,}{\,\,}{20}}
\hline\hline 
 & ,\mbox{Preselection CR} & ,$W$\mbox{+jets CR} & ,\mbox{four-jet }$\ttbar$\mbox{CR} & ,\mbox{five-jet }$\ttbar$\mbox{CR} \\
\hline
$t\bar{t}$ & ,50000\pm 4700 & ,2000\pm 400 &,19000\pm 600&,2100\pm 200 \\
$W$+jets & ,46000\pm 14000 & ,7000\pm 2900 &,3800\pm 800&,360\pm 170\\
Total & ,116000\pm 21000 &,12000\pm 3600&,26000\pm 1300&,2900\pm 440\\ \hline
Observed & ,110933 &,11858&,26197&,2736\\ \hline
300 \gev $W'$ & ,13900\pm 670&,930\pm 110&,3000\pm 400&,400\pm 80 \\
400 \gev $\phi$ & ,6100\pm 200&,430\pm 60&,1100\pm 100&,200\pm 20 \\
 \hline\hline 
 \end{tabular} 
 \end{table*}

Figure~\ref{fig:dataSR} shows the expected and observed $m_{tj}$ and $m_{\bar{t}j}$ distributions 
in the signal region. The data are found to be consistent with the SM expectation.
A variety of potential systematic effects are evaluated for the predicted signal and the background rates and shapes.
The dominant systematic effects of the jet energy scale~\cite{jets} and 
resolution~\cite{b-jer-unc} lead to uncertainties of up to 10\% on the total background rate and up to 21\% on 
the total 
signal expectation, depending on the mass of the new particle. The other dominant systematic uncertainty 
from the difference in $b$-tagging efficiency between simulation and data
leads to uncertainties of roughly 16\% on both the signal and background rates.
Effects due to lepton trigger uncertainties and 
ID efficiency as well as the energy scale and resolution are assessed using $Z\rightarrow ee$ and 
$Z\rightarrow \mu \mu$ data, which lead to systematic uncertainties of a few percent. 
Other potential systematic effects considered are the size of the small multi-jet background (assigned 100\% uncertainty); 
$\ttbar$ generator uncertainties (evaluated by comparing different results using the MC@NLO and POWHEG generators, 1--10\%); 
$\ttbar$ showering and fragmentation uncertainties (evaluated by comparing samples using both PYTHIA and HERWIG, 1--6\%); 
an uncertainty on the total integrated luminosity (3.9\%)~\cite{b-lumi,b-lumi2}; 
and the amount of QCD radiation for the signal and the $\ttbar$ background (approximately 10\%, evaluated using AcerMC). 
Total cross section uncertainties of 10\% (55\%) 
are used for the $\ttbar$ ($W$+jets) backgrounds. 

\begin{figure}[h!tbp]
\subfigure[]{\includegraphics[width=\columnwidth]{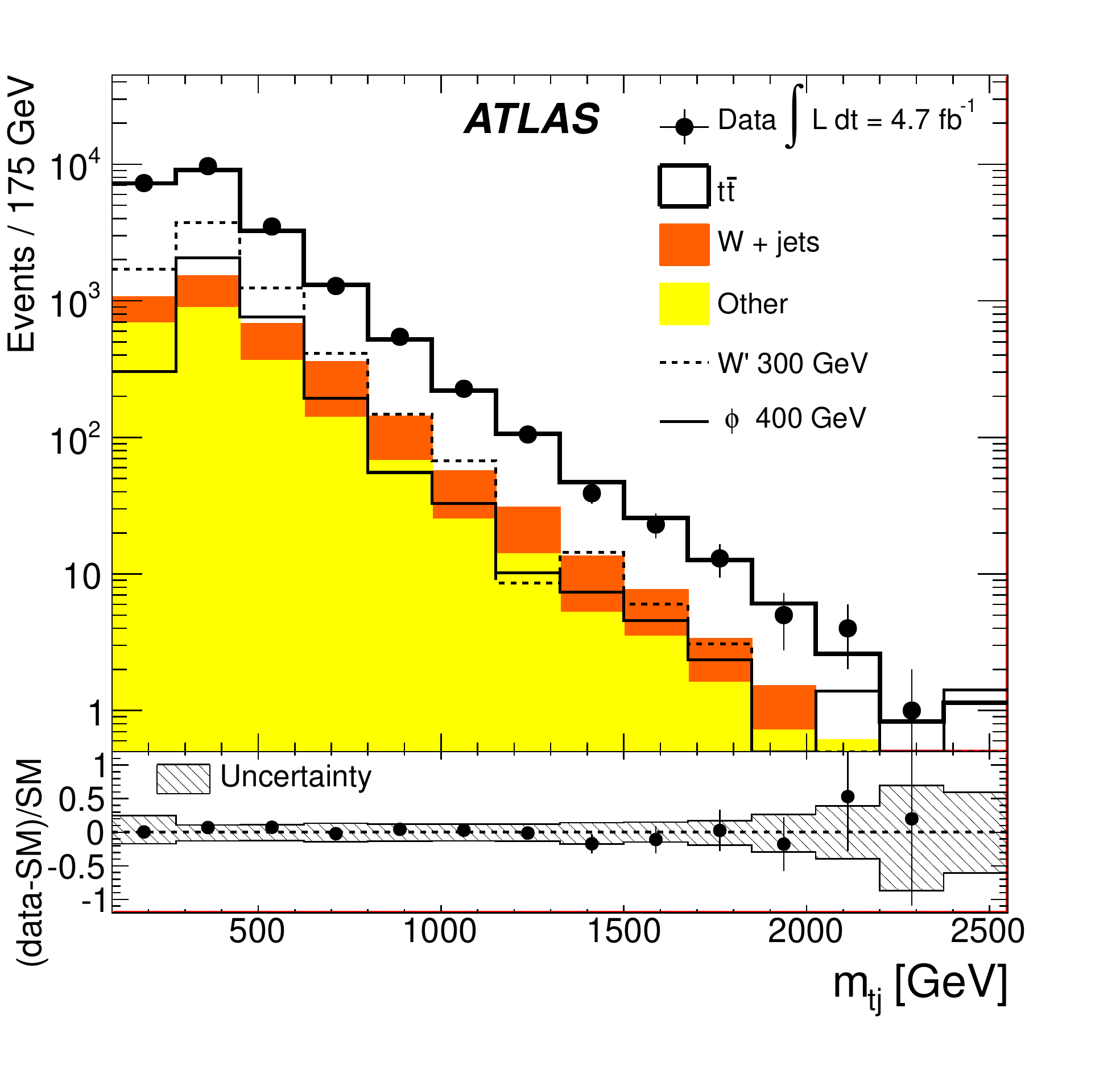}}
\subfigure[]{\includegraphics[width=\columnwidth]{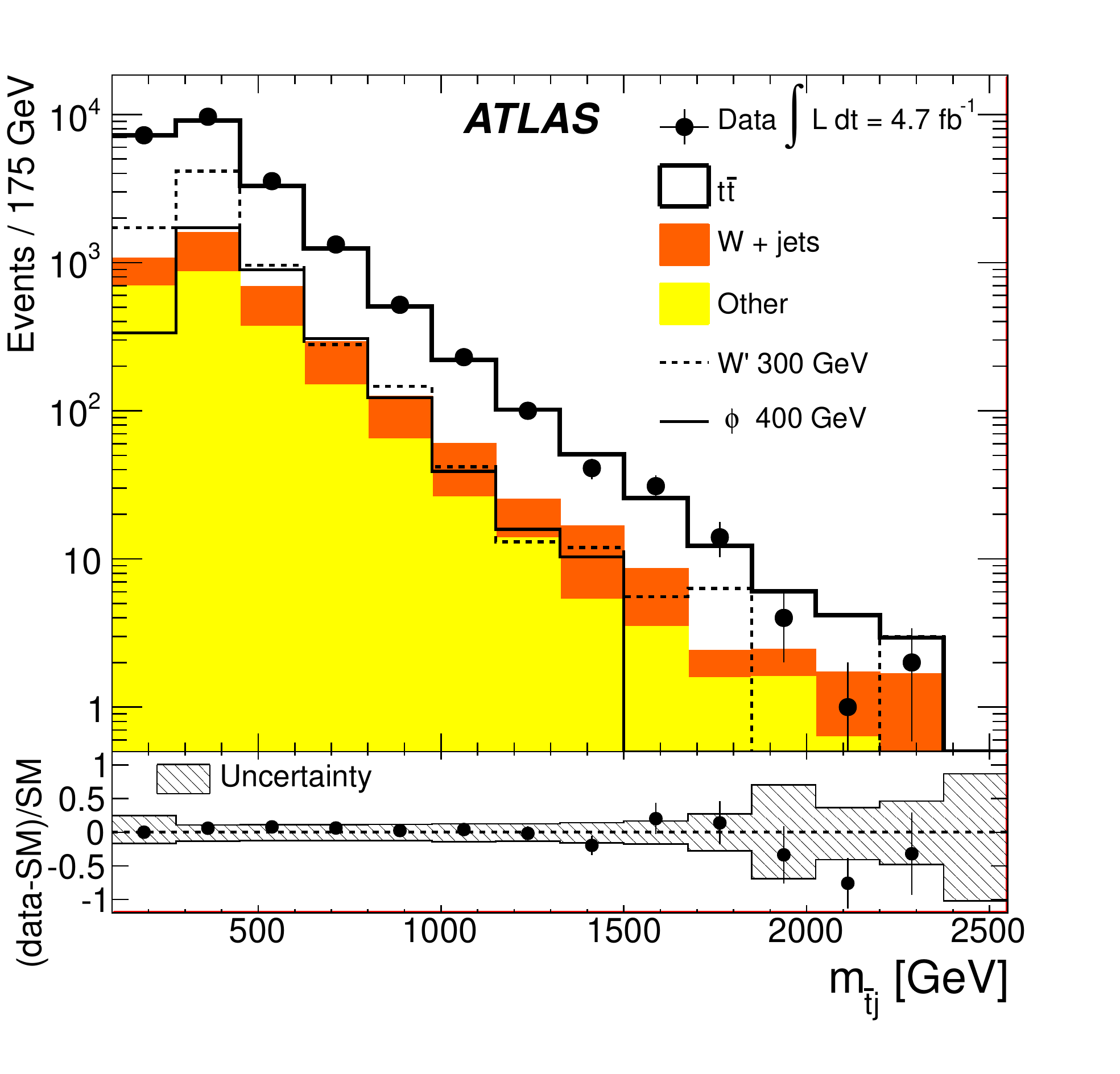}}
   \caption{Expected and observed distributions of (a) $m_{tj}$ and (b) $m_{\bar{t}j}$ in the signal region. 
The example signal distributions assume unit coupling for the new physics process.
The total uncertainty shown on the ratio includes both statistical and systematic effects. The ``other'' 
background category includes single top production, diboson production and multi-jet events.
  \label{fig:dataSR}
  }
\end{figure}

\begin{table*} 
  \caption{Expected and observed yields in different signal regions. The errors include all systematic 
uncertainties. Total refers to the total expected background, including $\ttbar$, $W$+jets and the 
other smaller backgrounds: single top production, diboson production and multi-jet events, which
are not tabulated separately here. 
Signal window eff. refers to the efficiency for the signal to fall inside 
the optimized two-dimensional mass window. The signal region yield is calculated  
in the mass window at each benchmark signal point. Signal $\sigma$ refers to the total expected signal cross section,
not taking into account the $t$ (or $\bar{t}$) plus jet branching fraction. }
 \label{tab:SR}
\begin{tabular}{l|D{,}{\,\,}{18}|D{,}{\,\,}{15}|D{,}{\,\,}{15}|D{,}{\,\,}{15}|D{,}{\,\,}{15}|}
\hline\hline 
& ,\mbox{Entire signal region} & ,\mbox{300~\gev $W'$} & ,\mbox{600~\gev $W'$} & ,\mbox{400~\gev$\phi$} & ,\mbox{800~\gev$\phi$} \\ \hline
\mbox{$m_{tj}$ window [GeV]} & & ,\mbox{344 $< m_{tj} <$ 494} & ,\mbox{566 $< m_{tj} <$ 904}& ,\mbox{401 $< m_{tj} <$ 455}& ,\mbox{766 $< m_{tj} <$ 819}\\
\mbox{$m_{\bar{t}j}$ window [GeV]} & & ,\mbox{292 $< m_{\bar{t}j} <$ 339}  & ,\mbox{549 $< m_{\bar{t}j} <$ 650} & ,\mbox{371 $< m_{\bar{t}j} <$ 608} & ,\mbox{628 $< m_{\bar{t}j} <$ 973} \\ 
Signal window eff. & & \multicolumn{1}{c|}{7.5\%} & \multicolumn{1}{c|}{9.9\%} & \multicolumn{1}{c|}{11.9\%} & \multicolumn{1}{c|}{5.7\%} \\\hline
$t\bar{t}$ & ,18000 \pm 3000 & ,740\pm 160 & ,270\pm 60&,660\pm 150&,60\pm10 \\
$W$+jets & ,1700 \pm 560 & ,60\pm30& ,30\pm 20 & ,80\pm 40 & ,8\pm5 \\
Total & ,22000\pm 3700 & ,820\pm 190 & ,320\pm 80&,780\pm 180&,70\pm20\\ \hline
Observed & ,22731 &,970&,343&,923&,77\\ \hline
Signal region yield & & ,560\pm 120&,98\pm24&,410\pm100& ,20\pm 6\\
\hline
Signal $\sigma$ & & \multicolumn{1}{c|}{19.0\,\mbox{pb}} & \multicolumn{1}{c|}{1.55\,\mbox{pb}} & \multicolumn{1}{c|}{7.9\,\mbox{pb}} & \multicolumn{1}{c|}{0.67\,\mbox{pb}} \\
 \hline\hline  
 \end{tabular} 
 \end{table*}

Expected and observed upper limits on the signal cross section are computed at discrete mass points as follows. 
For each benchmark signal mass point under consideration, a signal region is defined in the $m_{tj}$--$m_{\bar{t}j}$ plane.
When setting limits for the $W'$ ($\phi$) model, the $m_{tj}$ ($m_{\bar{t}j}$) window is significantly wider 
than the $m_{\bar{t}j}$ ($m_{tj}$) window to account for the 
fact that the resonance is predominantly in the $m_{\bar{t}j}$ ($m_{tj}$) system. The windows are optimized 
to maximize sensitivity, accounting for the full effect of systematic uncertainties. Typical mass windows
are shown in Table~\ref{tab:SR}. For each mass window, 
95\% confidence level (C.L.) upper limits on the signal cross section (times the branching ratio to $t$ or $\bar{t}$) are computed
using a single bin frequentist CL$_{\mbox{s}}$ method~\cite{b-cls}. 
No shape information is used within the mass windows.
Table~\ref{tab:SR} shows the expected and observed event yields in several of the signal region windows.
Expected and observed 95\% C.L. lower limits on the signal mass are derived, assuming a coupling of $g_R=1$ and $g_R=2$, 
and are shown in Figure~\ref{fig:limits}. Assuming that the cross section scales as $g_R^2$, the exclusion in the 
mass-coupling plane is shown in Figure~\ref{fig:limits2d}. As shown, most of the parameter space in this model, which was favored by
the Tevatron forward-backward asymmetry and cross section measurements~\cite{2dnote}, has been excluded.

\begin{figure}[h!tbp]
\centering
\subfigure[]{\includegraphics[width=\columnwidth]{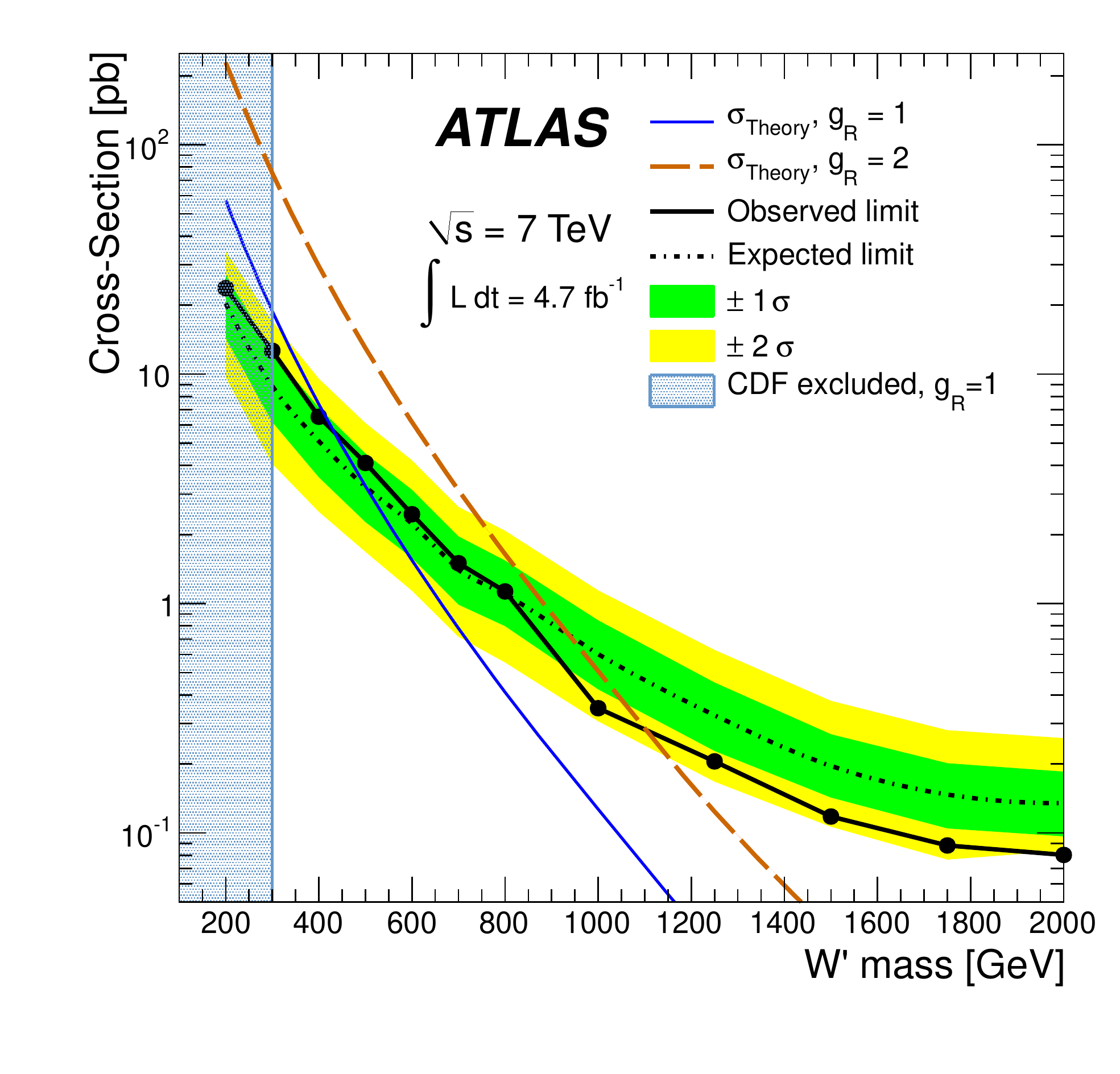}}
\subfigure[]{\includegraphics[width=\columnwidth]{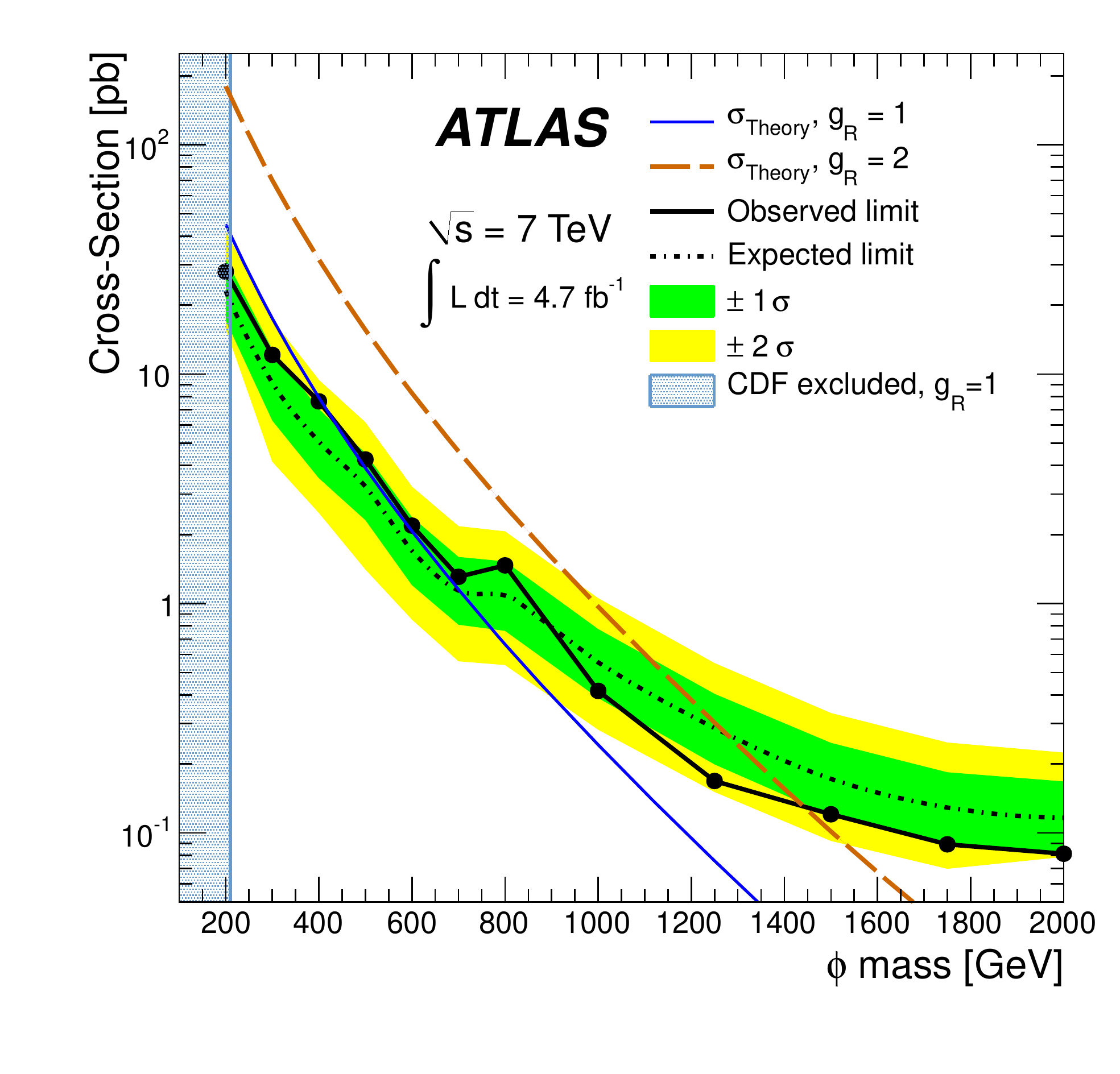}}
   \caption{Expected and observed 95\% C.L. upper limits on the (a) $W'$ and (b) $\phi$ model cross sections. 
   The CDF result is documented in Ref.~\cite{cdfsearch}. The $W'$ cross sections are NLO calculations, and the $\phi$ cross sections are LO calculations. 
  \label{fig:limits}
  }
\end{figure}

\begin{figure}[h!tbp]
\subfigure[]{\includegraphics[width=\columnwidth]{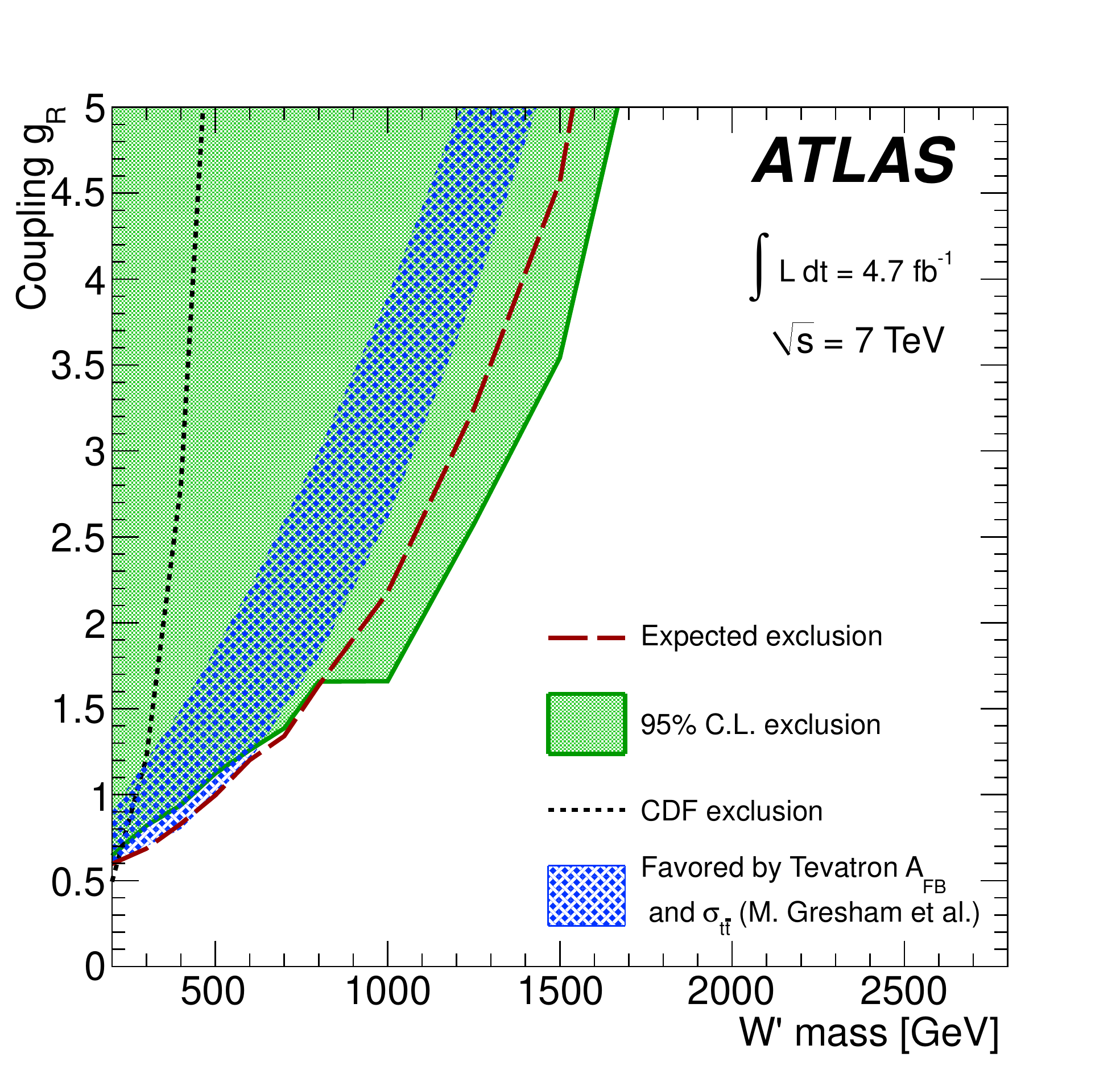}}
\subfigure[]{\includegraphics[width=\columnwidth]{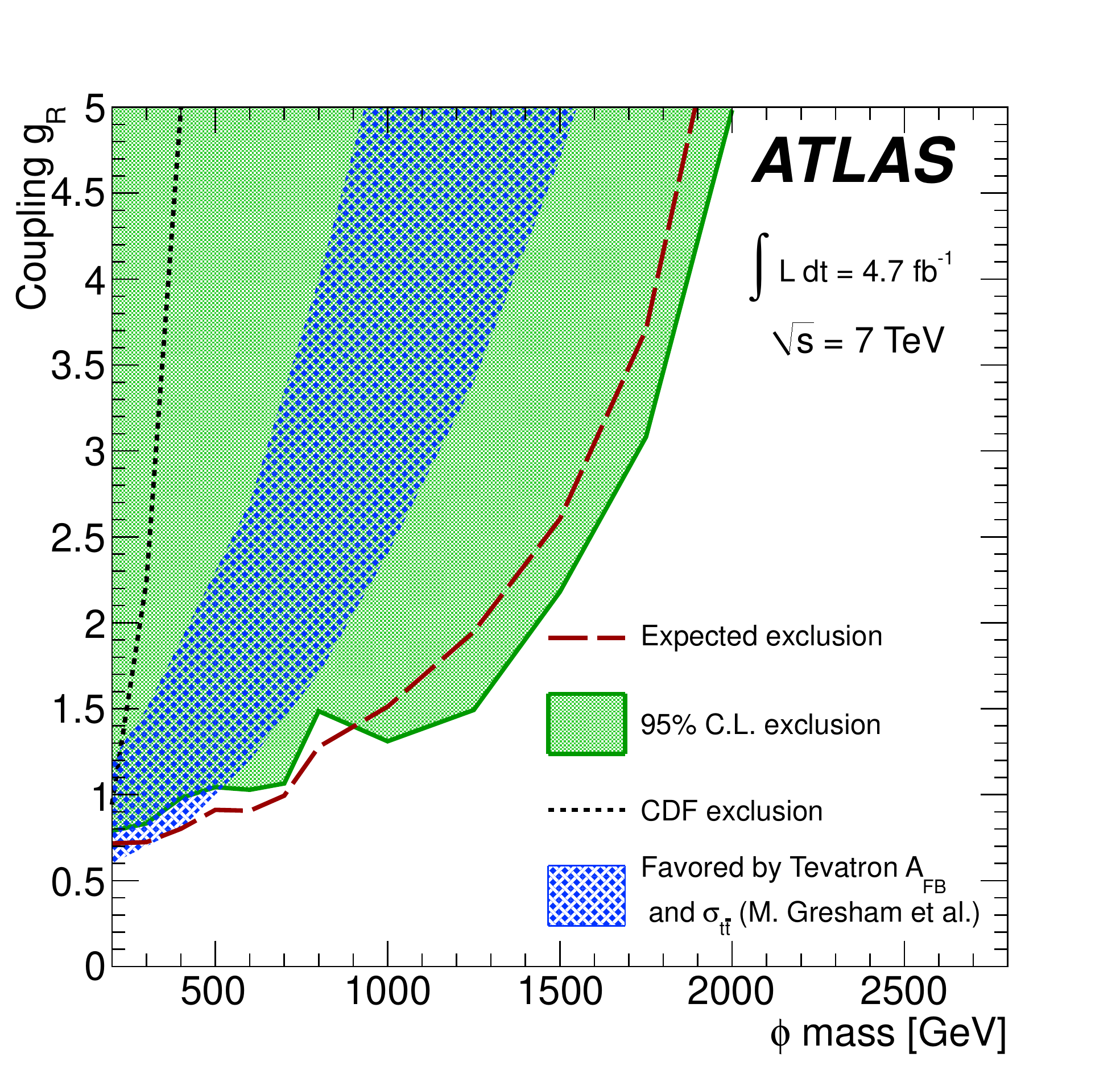}}
   \caption{Expected and observed 95\% C.L. upper limits on the (a) $W'$ and (b) $\phi$ model 
   cross sections assuming a cross section which scales with $g_R^2$. The hatched area shows the region of 
   parameter space excluded by this search at 95\% C.L.  
    The CDF result is documented in Ref.~\cite{cdfsearch}. The $W'$ cross sections are NLO calculations, and the $\phi$ cross sections are LO calculations. The region favored by the Tevatron $A_{FB}$ and $\sigma_{\ttbar}$ measurements is shown as the dark band~\cite{2dnote}.
  \label{fig:limits2d}
  }
\end{figure}

In conclusion, this paper presents a search for a new heavy particle $R$ in the $tj$ or $\bar{t}j$ system of 
$\ttbar$ plus extra jet events with the ATLAS detector. Such new particles have been proposed as a potential explanation of the difference from the SM values of the forward-backward asymmetries measured in top quark pair production at the Tevatron.
The full 2011 ATLAS $pp$ dataset (4.7 fb$^{-1}$) is used in the search.
Assuming unit coupling, the expected 95\% C.L. lower limit on the mass of the new particle is 500 (700)\,\gev 
in the $W'$ ($\phi$) model.
No significant excess of data above SM expectation is observed, and 95\% C.L. lower limits of 430\,\gev for both the $W'$ and $\phi$ models are set. At $g_R=2$, the limits are 1.10 (1.45)\,\tev for the $W'$ ($\phi$) model, with
expected limits of 0.93 (1.30)\,\tev. These are the most stringent limits to date on such models.
Most of the regions of parameter space for these models that are more consistent with the Tevatron forward-backward asymmetry and $\ttbar$ cross section measurements than the SM are excluded at 95\% C.L. by these results.

We thank CERN for the very successful operation of the LHC, as well as the
support staff from our institutions without whom ATLAS could not be
operated efficiently.

We acknowledge the support of ANPCyT, Argentina; YerPhI, Armenia; ARC,
Australia; BMWF and FWF, Austria; ANAS, Azerbaijan; SSTC, Belarus; CNPq and FAPESP,
Brazil; NSERC, NRC and CFI, Canada; CERN; CONICYT, Chile; CAS, MOST and NSFC,
China; COLCIENCIAS, Colombia; MSMT CR, MPO CR and VSC CR, Czech Republic;
DNRF, DNSRC and Lundbeck Foundation, Denmark; EPLANET and ERC, European Union;
IN2P3-CNRS, CEA-DSM/IRFU, France; GNSF, Georgia; BMBF, DFG, HGF, MPG and AvH
Foundation, Germany; GSRT, Greece; ISF, MINERVA, GIF, DIP and Benoziyo Center,
Israel; INFN, Italy; MEXT and JSPS, Japan; CNRST, Morocco; FOM and NWO,
Netherlands; BRF and RCN, Norway; MNiSW, Poland; GRICES and FCT, Portugal; MERYS
(MECTS), Romania; MES of Russia and ROSATOM, Russian Federation; JINR; MSTD,
Serbia; MSSR, Slovakia; ARRS and MVZT, Slovenia; DST/NRF, South Africa;
MICINN, Spain; SRC and Wallenberg Foundation, Sweden; SER, SNSF and Cantons of
Bern and Geneva, Switzerland; NSC, Taiwan; TAEK, Turkey; STFC, the Royal
Society and Leverhulme Trust, United Kingdom; DOE and NSF, United States of
America.

The crucial computing support from all WLCG partners is acknowledged
gratefully, in particular from CERN and the ATLAS Tier-1 facilities at
TRIUMF (Canada), NDGF (Denmark, Norway, Sweden), CC-IN2P3 (France),
KIT/GridKA (Germany), INFN-CNAF (Italy), NL-T1 (Netherlands), PIC (Spain),
ASGC (Taiwan), RAL (UK) and BNL (USA) and in the Tier-2 facilities
worldwide.

\clearpage


\onecolumngrid
\clearpage
\begin{flushleft}
{\Large The ATLAS Collaboration}

\bigskip

G.~Aad$^{\rm 48}$,
T.~Abajyan$^{\rm 21}$,
B.~Abbott$^{\rm 111}$,
J.~Abdallah$^{\rm 12}$,
S.~Abdel~Khalek$^{\rm 115}$,
A.A.~Abdelalim$^{\rm 49}$,
O.~Abdinov$^{\rm 11}$,
R.~Aben$^{\rm 105}$,
B.~Abi$^{\rm 112}$,
M.~Abolins$^{\rm 88}$
% ...
\bigskip

$^{1}$ School of Chemistry and Physics, University of Adelaide, Adelaide, Australia\\
% ...
$^{11}$ Institute of Physics, Azerbaijan Academy of Sciences, Baku, Azerbaijan\\
$^{12}$ Institut de F\'{i}sica d'Altes Energies and Departament de F\'{i}sica de la Universitat Aut\`{o}noma de Barcelona and ICREA, Barcelona, Spain\\
% ...
$^{21}$ Physikalisches Institut, University of Bonn, Bonn, Germany\\
% ...
$^{48}$ Fakult\"{a}t f\"{u}r Mathematik und Physik, Albert-Ludwigs-Universit\"{a}t, Freiburg, Germany\\
$^{49}$ Section de Physique, Universit\'{e} de Gen\`{e}ve, Geneva, Switzerland\\
% ...
$^{88}$ Department of Physics and Astronomy, Michigan State University, East Lansing MI, United States of America\\
% ...
$^{105}$ Nikhef National Institute for Subatomic Physics and University of Amsterdam, Amsterdam, Netherlands\\
% ...
$^{111}$ Homer L. Dodge Department of Physics and Astronomy, University of Oklahoma, Norman OK, United States of America\\
$^{112}$ Department of Physics, Oklahoma State University, Stillwater OK, United States of America\\
% ...
$^{115}$ LAL, Universit\'{e} Paris-Sud and CNRS/IN2P3, Orsay, France\\
% ...
$^{t}$ Also at Manhattan College, New York NY, United States of America\\
% ...
\end{flushleft}



\begin{thebibliography}{unsrt}
\bibitem{cmsfootnote} There are several differences between the models in Refs.~\cite{TFVResonances}
and~\cite{TFVResonances2}. The Lagrangian in the former (used in this paper) includes a factor of $1/\sqrt{2}$, and the one 
in the latter (used by CMS) does not. In addition, Ref.~\cite{TFVResonances2} includes additional non-resonant 
diagrams with cross section that scale as $g_R^4$. Such diagrams are not included in Ref.~\cite{TFVResonances}.
\bibitem{atlasfootnote}ATLAS uses a right-handed coordinate system with its origin at the nominal 
interaction point (IP) in the center of the detector and the $z$-axis along the beam pipe. The $x$-axis 
points from the IP to the center of the LHC ring, and the $y$-axis points upward. Cylindrical coordinates $(r,\phi)$ 
are used in the transverse ($x-y$) plane, $\phi$ being the azimuthal angle around the beam pipe. The pseudorapidity 
is defined in terms of the polar angle $\theta$ as $\eta=-\ln\tan(\theta/2)$.
\end{thebibliography}
\end{document}